\documentclass[prl,aps,twocolumn]{revtex4-1}
\usepackage{bm}
\usepackage{graphicx}
\usepackage{amsmath}
\usepackage{amssymb}
\usepackage[utf8]{inputenc}
\usepackage[T1]{fontenc}
\usepackage{color}
\usepackage{upgreek}
\usepackage{subfigure}
\usepackage[unicode=true,colorlinks=true,citecolor=blue]{hyperref}
\usepackage{placeins}
\usepackage{lipsum}
\usepackage{epsfig}
\usepackage[utf8]{inputenc}
\usepackage{wrapfig}
\newcommand{\nix}[1]{}
\usepackage[version=4]{mhchem}
\usepackage{siunitx}

\newcommand{\HgTe}{\ce{HgTe}~}
\newcommand{\HgCdTe}{\ce{HgCdTe}~}

\newcommand{\CdTe}{\ce{CdTe}~}

\newcommand{\kWcm}[1]{\SI{#1}{\kilo\watt\per\centi\meter\squared}}
\newcommand{\MilliMeter}[1]{\SI{#1}{\milli\metre}}
\newcommand{\MilliElectronVolt}[1]{\SI{#1}{\milli\electronvolt}}
\newcommand{\MilliMeterSquared}[1]{\SI{#1}{\milli\metre\squared}}
\newcommand{\MicroMeter}[1]{\SI{#1}{\micro\meter}}
\newcommand{\NanoMeter}[1]{\SI{#1}{\nano\meter}}
\newcommand{\NanoAmpere}[1]{\SI{#1}{\nano\ampere}}

\newcommand{\CentiMeterSquaredPerVoltSeconds}[1]{\SI{#1}{\centi\meter\squared\per{\volt\second}}}

\newcommand{\TeraHertz}[1]{\SI{#1}{\tera\hertz}}
\newcommand{\NanoSecond}[1]{\SI{#1}{\nano\second}}

\begin{document}

\title{Impact ionization induced by terahertz radiation in HgTe quantum wells of critical thickness}

\author{S. Hubmann$^1$, G.V. Budkin$^2$, M. Urban$^1$, V.V. Bel'kov$^2$, A.P.~Dmitriev$^2$, J. Ziegler$^1$, D.A. Kozlov$^3$, N.N. Mikhailov$^3$, S.A. Dvoretsky$^3$, Z.D. Kvon$^3$, D. Weiss$^1$ and S.D. Ganichev$^1$}

\affiliation{$^1$Terahertz Center, University of Regensburg, 93040 Regensburg, Germany}

\affiliation{$^2$Ioffe Institute, 194021 St. Petersburg, Russia}

\affiliation{$^3$Rzhanov Institute of Semiconductor Physics, 630090 Novosibirsk, Russia}


\begin{abstract}
 We report on the observation of terahertz (THz) radiation induced band-to-band impact ionization in \HgTe quantum well (QW) structures of critical thickness, which are characterized by a nearly linear energy dispersion. The THz electric field drives the carriers initializing electron-hole pair generation. The carrier multiplication is observed for photon energies less than the energy gap under the condition that the product of the radiation angular frequency $\omega$ and momentum relaxation time $\tau_{\text l}$ larger than unity. In this case,  the charge carriers acquire high energies solely because of collisions in the presence of a high-frequency  electric field. The developed microscopic theory shows that the probability of the light impact ionization is proportional to $\exp(-E_0^2/E^2)$, with the radiation electric field amplitude $E$ and the characteristic field parameter $E_0$. As  observed in experiment, it exhibits a strong frequency dependence for $\omega \tau \gg 1$ characterized by the characteristic field $E_0$ linearly increasing with the radiation frequency $\omega$. 
\end{abstract}

\maketitle

\section{Introduction}
\label{introduction}

Impact ionization across the band edges and its inverted process -  Auger recombination - as well as impact ionization of  impurities  are the most important autocatalytic processes in semiconductors. 
They have been studied extensively not only because of fundamental interest in these nonlinear phenomena, but also due to their great practical importance for IMPATT diodes (impact ionization avalanche transit time)~\cite{Sze2006}, high efficiency solar cells~\cite{Landsberg1993}, and photodetectors with internal amplification like avalanche photodiodes, particularly useful in the case of fiber-optic communication systems ~\cite{Capasso1985}.
Aside from being excited by a $dc$ electric field like the aforementioned processes impact ionization can also be excited by the $ac$ electric field of THz radiation. Such a process has been observed first in bulk \ce{InSb} crystals and was termed \emph{light} impact ionization~\cite{Ganichev1984,Ganichev1986a}. With the development of high-power THz laser systems like molecular lasers, free-electron lasers, and Ti:Sapphire based THz-systems, there has been a steady increase on experimental and theoretical research interest in the field of THz radiation-induced impact ionization, carrier multiplication, and nonperturbative nonlinearities in  three- and two-dimensional semiconductor systems ~\cite{Ganichev1994,Markelz1996,Gaal2006,Cao2003,Wen2008,Hoffmann2009,Kuehn2010,Watanabe2011,Hirori2011,Ho2011,Shinokita2013,Lange2014,Vampa2015,Hafez2016,Tarekegne2017,Vicario2017,Chai2018,Asmontas2018,Hubmann2019}, for reviews see~\cite{Ganichev2005,Hirori2016}.
Recently it has been shown that impact ionization and Auger recombination processes can also be efficiently excited and probed by THz radiation in graphene~\cite{Strait2011,Tani2012,Tielrooij2013,Mittendorff2014,Oladyshkin2017}.
Interband carrier–carrier scattering such as impact ionization- and/or Auger-type processes in graphene are of particular importance. Because of the peculiar linear dispersion in Dirac materials and the conservation laws they are allowed only when the momenta of all the particles involved in the ionization/recombination are co-linear. However, it has been shown, that these processes becomes non-vanishing either due to the extent of a small difference between the carrier dispersion from the linear one (e.g. trigonal warping) or due to many body effects such as plasmon-assisted processes or additional scattering by an impurity or phonon~\cite{Golub2011a,Kotov2012,Pirro2012,Brida2013,Tomadin2013,Alymov2018}. A suppression of the Auger recombination has also been addressed for \ce{HgTe}-based QW structures with symmetric dispersion laws in conduction and valence bands~\cite{Morozov2017}. This suppression has been used to obtain band-band population inversion and stimulated THz emission, which, because of the efficient nonradiative Auger recombination, can not be achieved  in conventional narrow band semiconductors. When approaching the critical thickness in \ce{HgTe}/\HgCdTe QWs, the band structure gets almost linear~\cite{Bernevig2006}, i.e., similar to that of graphene, but characterized by electron spin instead of pseudo-spin. The linear dispersion in \ce{HgTe}/\HgCdTe QWs with critical thickness has been demonstrated in transport~\cite{Buettner2011} and THz experiments~\cite{Kvon2011,Olbrich2013,Shuvaev2016}. 

Here we show that excitation of such QWs by intense THz radiation results in an efficient impact ionization process. Applying monochromatic radiation with frequencies from 0.6 to \TeraHertz{2} we observed a photoconductivity signal rising superlinearly with the radiation intensity $I$. The photoconductivity is caused by the generation of electron-hole pairs and, in a large range of radiation intensities, varies as $\exp(-E_0^2/E^2)$, where $E$ is the radiation electric field amplitude $E\propto \sqrt I$ and $E_0$ is the characteristic field parameter. Furthermore, it shows a strong frequency dependence decreasing with the frequency increase. The observed field and frequency dependencies indicate that the generation of electron-hole pairs is caused by the light impact ionization in high frequency electric fields. As shown in Ref.~\cite{Keldysh1965}, light impact ionization is divided into two regimes: (i) the quasi-static, in which the angular radiation frequency $\omega=2\pi f$ is much lower than the reciprocal momentum relaxation time $\tau^{-1}$ and the ionization takes place within a half period of the field, and (ii) the high-frequency regime with $\omega\gg\tau^{-1}$, in which carriers acquire the ionization energy due to collisions. In our experiments $\omega \gtrsim \tau^{-1}$ which corresponds to the latter regime characterized by strong frequency dependence. We developed a theory considering impact ionization for the real band structure of HgTe QWs with thickness close to the critical one. The theory describes both the quasi-static and high-frequency regimes. It describes all experimental findings well and shows that the observed nonlinear photoconductivity is caused by the latter regime also known as \emph{light} impact ionization.

\section{Samples and methods}
\label{samples_methods}

\begin{figure}
	\centering
	\includegraphics[width=\linewidth]{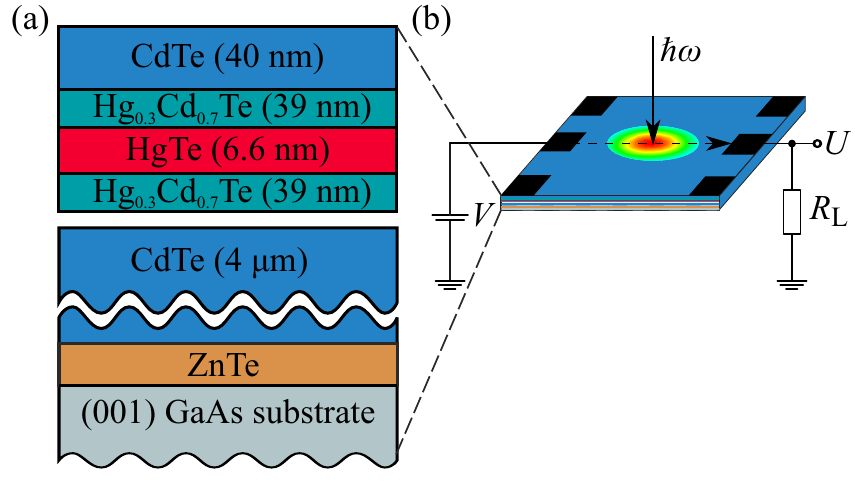}
	\caption{Panel (a): Structure composition. Panel (b): Sketch of the setup used for the photoconductivity measurements. 
	}
	\label{setup}
\end{figure}

\begin{figure}
	\centering
	\includegraphics[width=\linewidth]{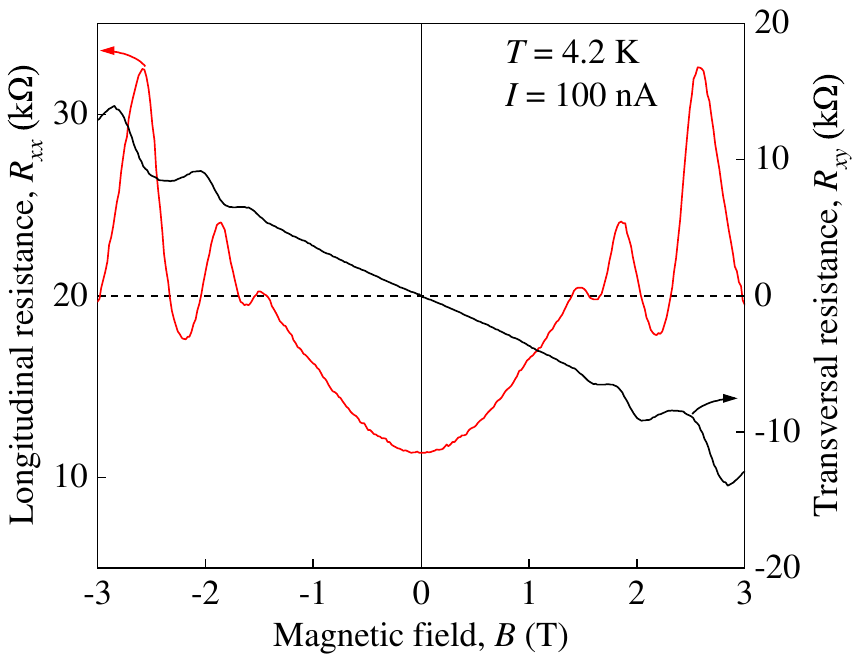}
	\caption{Magnetotransport data obtained at liquid helium temperature with a current of \NanoAmpere{100}. The red line shows the longitudinal resistance $R_{xx}$ while the black line shows the transversal resistance $R_{xy}$.}
	\label{transport}
\end{figure}

The samples studied in this work are \ce{HgTe}/HgCdTe QWs grown by molecular beam epitaxy on (001)-oriented GaAs substrates by an analogous procedure as described in \cite{Dvoretsky2010}. The \NanoMeter{6.6} wide QWs were surrounded by two \NanoMeter{39} thick \ce{Hg_{0.3}Cd_{0.7}Te}~ barriers, see Fig.~\ref{setup}(a). In order to relax strain stemming from the lattice mismatch between the GaAs substrate and \HgCdTe a \MicroMeter{4} thick \CdTe buffer layer was grown in between. This structure composition leads to a almost linear energy spectrum \cite{Bernevig2006,Buettner2011,Kvon2011,Olbrich2013}. Our $\bm k\cdot\bm p$ calculations presented below show that in the structure 
a small band gap of about $\varepsilon_{\text g}=\MilliElectronVolt{4.5}$ should be present. 
The sample size was $5\times\MilliMeterSquared{5}$ in a van-der-Pauw sample geometry, see Fig.~\ref{setup}(b). Ohmic contacts were fabricated by indium soldering to make photoconductivity and magnetotransport 
measurements possible. Applying magnetotransport measurement we obtained a carrier density of \SI{1.7e11}{\per\centi\meter\squared} and a mobility of \CentiMeterSquaredPerVoltSeconds{5700}, see Fig.~\ref{transport}.

To apply a high-frequency electric field we used the THz radiation from an optically pumped high-power pulsed molecular gas THz laser \cite{Ganichev1995,Olbrich2011,Drexler2012}. This laser system features several frequency lines between 0.6 and \TeraHertz{3.3} (2.5 to \MilliElectronVolt{14}) with a pulse duration of about \NanoSecond{100}, a repetition rate of \SI1{\hertz} and a gaussian beam shape. The latter has been measured by a pyroelectric camera \cite{Dantscher2015}. Using a parabolic mirror the beam was focused onto the sample with a spot diameter of about \MilliMeter{2.5}. The time structure of THz pulses was controlled by a fast room temperature photon-drag detector~\cite{Ganichev1985}. The samples were placed in an optical temperature-regulated continuous flow cryostat with $z$-cut crystal quartz windows. The measurements have been carried out in a temperature range from 4.2 to \SI{90}{\kelvin}. All experiments were performed illuminating the sample with the THz laser radiation under normal incidence. In order to vary the laser radiation intensity a crossed polarizer setup was used: First the linearly polarized radiation passed a wire grating polarizer, which was rotated to modify the radiation intensity. Then a second polarizer at a fixed position ensured a fixed output polarization.

The $dc$ photoconductivity was measured using the setup shown in Fig.~\ref{setup}(b). A $dc$ bias voltage $V=\SI{0.3}{\volt}$ was applied and the voltage drop $U$ in response to the laser pulse was measured across a load resistor $R_{\text L}$. 
The photoconductivity signal can be separated from possible photocurrent contributions by subtracting signals detected for negative and positive bias voltages and dividing by 2, since the photocurrent contributions are not sensitive to the bias voltage in contrast to the photoconductivity signal.


\section{Results}
\label{results}

First, we discuss the data for radiation with photon energies $\hbar\omega$ smaller than the band gap $\varepsilon_{\text g}\approx \MilliElectronVolt{4.5}$. 

\begin{figure}
	\centering
	\includegraphics[width=\linewidth]{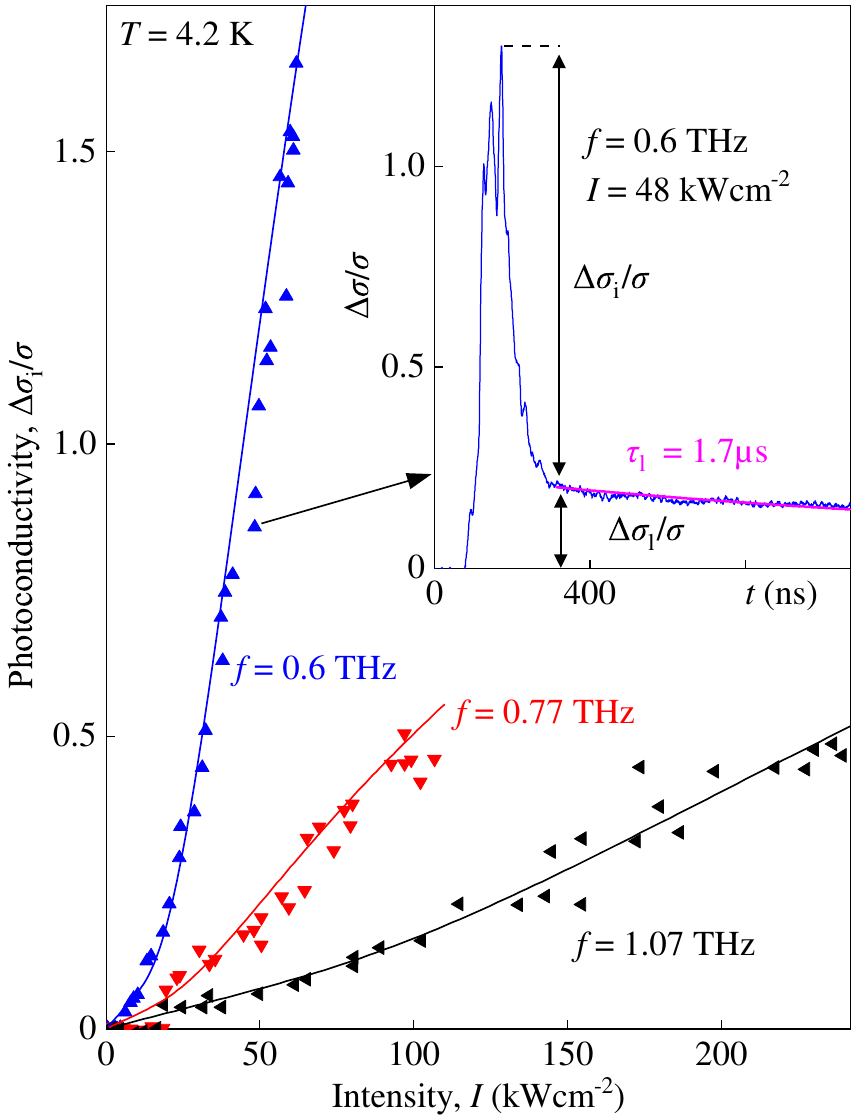}
	\caption{Intensity dependence of the photoconductivity signal $\Delta\sigma_{\text i}/\sigma$ for frequencies of 0.6 (blue triangles), 0.77 (red triangles), and \TeraHertz{1.07} (black triangles). Solid lines present the fit according to Eq.~\eqref{Esquare} with the fitting parameters $A$, $B$, and $I_0$. Note that for all frequencies $\hbar\omega<\varepsilon_{\text g}$ is valid. Inset shows the typical kinetics of the photoconductivity pulse for a frequency of \TeraHertz{0.6} and an intensity of \kWcm{48}. The magenta line shows a fit of the exponential decay according to $\Delta\sigma_{\text l}/\sigma\propto \exp\left(-t/\tau_{\text l} \right)$}.
	\label{intensity}
\end{figure}

The inset in Fig.~\ref{intensity} shows the photoconductive response observed applying radiation with a frequency of $f=\TeraHertz{0.6}$ to the sample cooled down to liquid helium temperature. The detected signal temporal shape, being characteristic for all  frequencies 
and all studied temperatures, consists of two parts characterized by different response times and relative amplitudes.  The first part  ($\Delta\sigma_{\text i}/\sigma$)  has a response time in the range of several tens of nanoseconds. The second part ($\Delta\sigma_{\text l}/\sigma$) has a substantially longer response time, being in the microsecond range. While at high intensities the contribution $\Delta\sigma_{\text i}/\sigma$ yields the highest signal at low intensities the situation changes and $\Delta\sigma_{\text l}/\sigma$ dominates.  In the following we will first focus on  $\Delta\sigma_{\text i}/\sigma$, which, as we show below, is caused by light impact ionization. Due to substantial difference in the signal kinetic this contribution can be easily extracted from the total signal. 

\begin{figure}
	\centering
	\includegraphics[width=\linewidth]{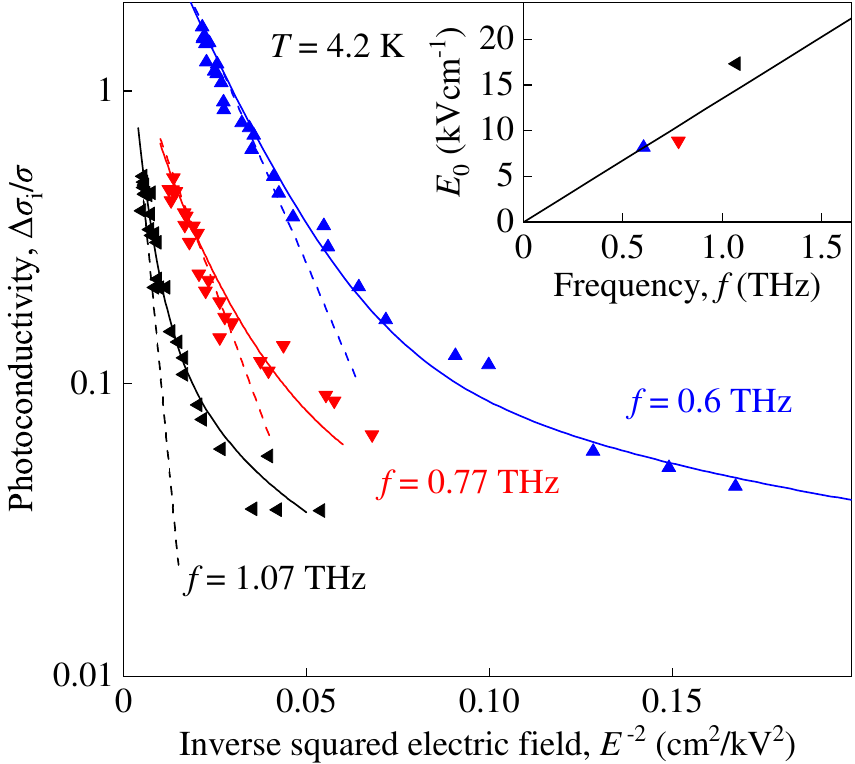}
	\caption{Dependency of $\Delta\sigma_{\text i}/\sigma$ on the inverse squared electric field  obtained for frequencies of 0.6 (blue triangles), 0.77 (red triangles), and \TeraHertz{1.07} (black triangles). The data are presented in a half-logarithmic plot. Solid lines show the fits according to Eq.~\eqref{Esquare}. Dashed lines show fits after exponential term in right hand side of Eq.~\eqref{Esquare} which gives the same field dependence as the theoretical Eq.~\eqref{generation_rate_proportionality}. Inset shows the dependence of the fit parameter $E_0$ on frequency. Solid line is a linear fit after Eq.~\eqref{generation_rate_proportionality}.}
	\label{field}
\end{figure}

The intensity dependencies of $\Delta\sigma_{\text i}/\sigma$ for three frequencies and $T=\SI{4.2}{\kelvin}$ are shown in Fig.~\ref{intensity}. The data demonstrate a superlinear dependence of the signal on the radiation intensity. The data are fitted well by
\begin{equation}
\label{Esquare}
\frac{\Delta\sigma_{\text i}}{\sigma}=A\cdot I+B\cdot\exp\left(-\frac{I_0}{I}\right)=A_{\text E}\cdot E^2+B\cdot\exp\left(-\frac{E_0^2}{E^2}\right)\:,
\end{equation}
Here, $I=(E^2\cdot n_{\omega}/(2Z_0)$ is the radiation intensity, $E$ is the radiation electric field, $n_{\omega}$ is the refractive index, $Z_0$ is the vacuum impedance, $I_0=(E^2_0\cdot n_{\omega}/(2Z_0)$ is the characteristic intensity, $E_0$ is the characteristic electric field, and $A_{\text E}=A\cdot n_{\omega}/(2Z_0)$. The fits shown in Fig.~\ref{intensity} are obtained using  fitting parameters $I_0$, $A$, and $B$. An important observation is that, the nonlinearity is defined by the characteristic intensity $I_0\propto E_0^2$ and, that lowering the radiation frequency results in a substantial increase of the signal amplitude. As we show below, the exponential part of the right-hand side in Eq.~\eqref{Esquare} with $E_0\propto\omega$ coincides with that obtained from the theoretical examination of light-impact ionization, see Eq.~\eqref{generation_rate_proportionality}.  Re-plotting the data in a half-logarithmic plot as a function of the inverse squared electric field $E^{-2}$, Fig.~\ref{field}, we obtain that at high radiation electric fields the exponential term 
describes well our results. As it is shown in the inset in Fig.~\ref{field}, scaling of $E_0$ with $\omega$, being characteristic for the light impact ionization \cite{Ganichev1984,Hubmann2019},  is also detected. At low intensities, however, we find a deviation from this behavior and the signal is determined by the first term in the left-hand side of Eq.~\eqref{intensity}.


\begin{figure}
	\centering
	\includegraphics[width=\linewidth]{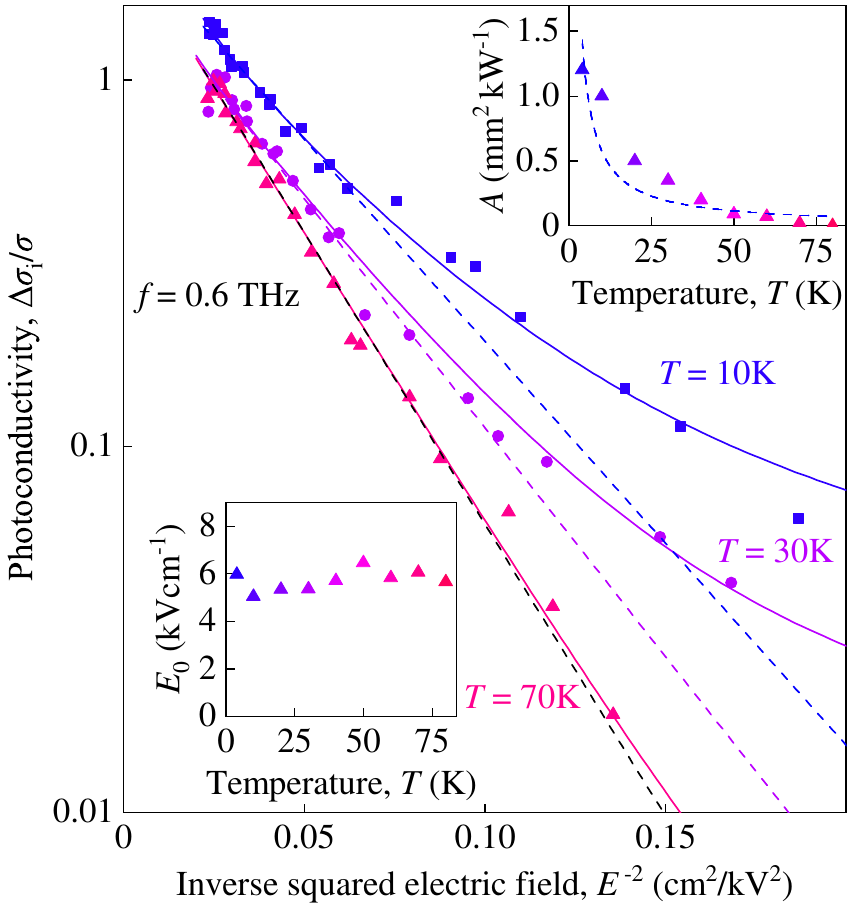}
	\caption{Photoconductivity $\Delta\sigma_{\text i}/\sigma$ as a function of the inverse squared electric field $E^{-2}$  measured for three temperatures:  \SI{10}{\kelvin} (blue squares), \SI{30}{\kelvin} (purple circles) and \SI{70}{\kelvin} (pink triangles). The data are obtained at \TeraHertz{0.6} and are presented in a half-logarithmic plot. Solid lines show fits according to Eq.~\eqref{Esquare}. Dashed lines show fits according to the exponential term in right hand side of Eq.~\eqref{Esquare} which gives the same field dependence as the theoretical Eq.~\eqref{generation_rate_proportionality}. Insets show temperature dependencies of $E_0^2$ and parameter $A$. 
	}
	\label{temperature}
\end{figure}

Rising temperature results in a slight increase of the characteristic electric field $E_0$ and a substantial decrease of the fitting parameter $A$ defining the first term in Eq.~\eqref{Esquare}.  Figure~\ref{temperature}  shows the data obtained for three different temperatures at radiation frequency $f=\TeraHertz{0.6}$ indicating that already at $T=\SI{70}{\kelvin}$ the exponential term in Eq.~\eqref{Esquare} dominates the photoconductivity in the whole range of the radiation intensity. Temperature evolution of $E_0^2$ and parameter $A$ are shown in the insets of Fig.~\ref{temperature}.

\begin{figure}
	\centering
	\includegraphics[width=\linewidth]{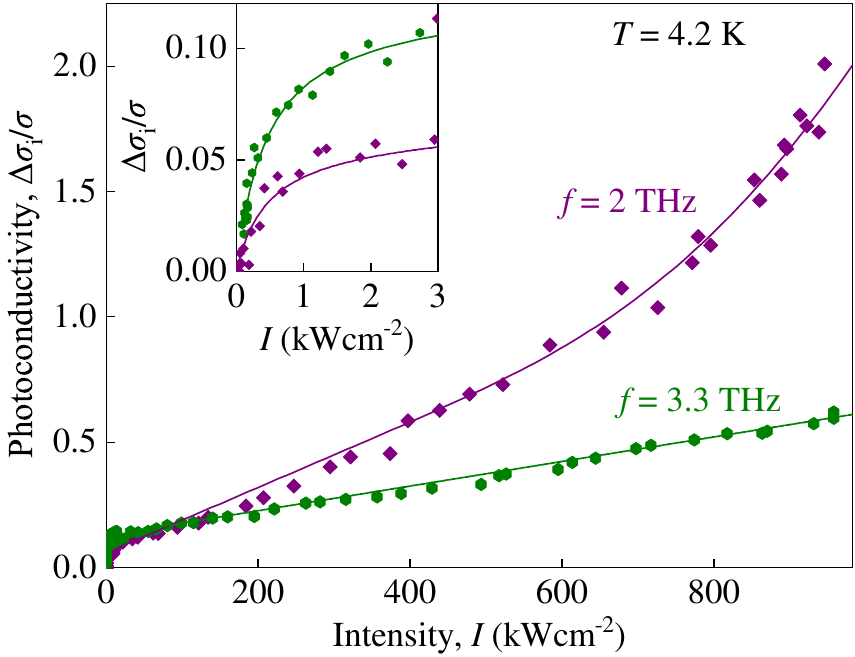}
	\caption{ Intensity dependences of $\Delta\sigma_{\text i}/\sigma$ at 2 (violet diamonds) and \TeraHertz{3.3} (green hexagons). Solid lines present fits according to Eq.~\eqref{highfreqfit}. Note that for both frequencies $\hbar\omega>\varepsilon_{\text g}$ is valid. The inset shows a zoom-in for low intensities.}
	\label{highfrequency}
\end{figure}

Now we discuss the photoconductivity  obtained for photon energies higher than the energy gap, $\hbar\omega>\varepsilon_{\text g}$. An increase of the photon energy qualitatively changes the intensity dependence of $\Delta\sigma_{\text i}/\sigma$: At highest frequency used ($f=\TeraHertz{3.3}$), we observed instead of superlinear behavior that the signal saturates with rising intensity, see green triangles and line in Fig.~\ref{highfrequency}. Saturation with increasing radiation intensity is also clearly seen for $f=\TeraHertz{2}$, however, for this frequencies  a superlinear behavior shows up and becomes dominant at high intensities, see violet diamonds and line Fig.~\ref{highfrequency}. Our experiments show that for  $\hbar\omega>\varepsilon_{\text g}$ an additional term describing the saturation of the photoconductivity  should be added to the Eq.~\eqref{Esquare}  and the  overall intensity dependence is given by 
\begin{equation}
\label{highfreqfit}
\frac{\Delta\sigma_{\text i}}{\sigma}=C\frac{I}{1+I/I_S}+A\cdot I+B\cdot\exp\left(-\frac{I_0}{I}\right)
\end{equation}
with the prefactor 
$C$ and the saturation intensity $I_S$ used to describe the signal saturation.

\begin{figure}
	\centering
	\includegraphics[width=\linewidth]{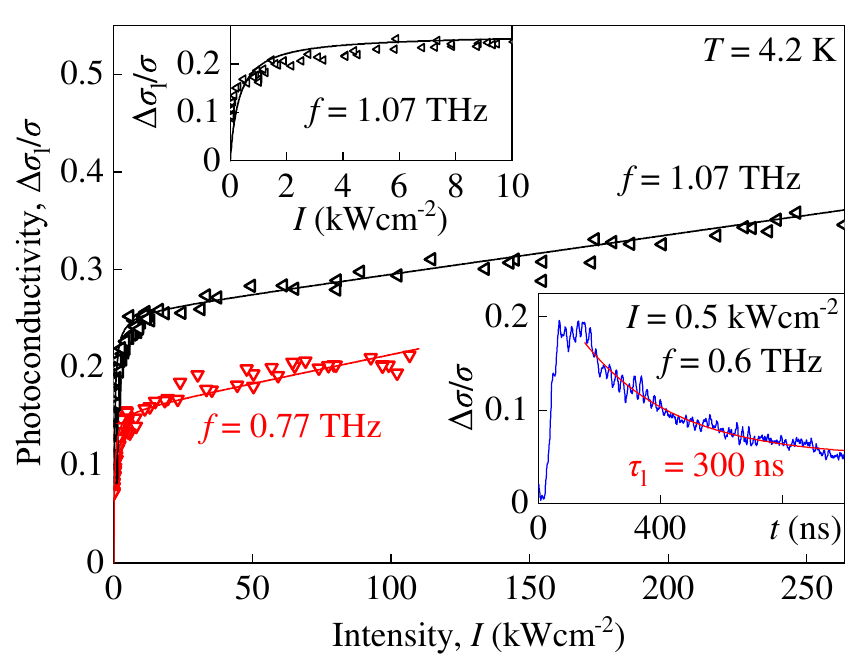}
	\caption{Intensity dependence of the slow component of the photoconductivity signal $\Delta\sigma_{\text l}/\sigma$. The data are shown for frequencies of 0.77 (red triangles) and \TeraHertz{1.07} (black triangles). Solid lines present fits according to Eq.~\eqref{linearfit}. The left inset shows a zoom-in for low intensities. The right inset shows the typical kinetics of the photoconductivity pulses obtained for $f= \TeraHertz{0.6}$ and 
	$I=\kWcm{0.5}$. The red line shows an exponential decay fit according to $\Delta\sigma/\sigma\propto \exp\left(-t/\tau_{\text l} \right)$}.
	\label{slow}
\end{figure}

Finally we describe the slow photoconductive signal component $\Delta\sigma_{\text l}/\sigma$. For the temperature of \SI4{\kelvin} and at low intensities the slow response dominates the signal and is characterized by a time constant of $\SI{0.3}{\micro\second}$, see inset in Fig.~\ref{slow}. Figure~\ref{slow} shows the corresponding intensity dependence for frequencies 0.77 and \TeraHertz{1.07}. The data reveal that the slow component of the signal at low intensity increases linearly with $I$ and saturates at high intensities. The data are fitted according to
\begin{equation}
\label{linearfit}
\frac{\Delta\sigma_{\text l}}{\sigma}=C_l\frac{I}{1+I/I_{l,S}}+A_l\cdot I
\end{equation}
with the fitting parameters $A_l$, $I_{l,S}$, and $C_l$. Because of the saturation, at high intensity the time dynamics of the signal gets substantially slower reaching time constants of about \SI{2}{\micro\second}, see inset in Fig.~\ref{intensity}. Rising the temperature $\Delta\sigma_{\text l}/\sigma$ vanishes for $T>\SI{20}{\kelvin}$. (not shown) This fact together with the slow kinetic of the photoconductive signal indicates that it is caused by  ionization of impurities in the \HgTe QW. Indeed, at high temperatures impurities get thermally ionized and consequently, the photosignal vanishes. Furthermore, at high radiation intensities impurities become completely ionized resulting in the signal saturation as detected in our experiment. Extrinsic photoconductivity and its saturation are well known processes and, therefore, their consideration is out of scope of our paper.


\begin{figure}
	\centering
	\includegraphics[width=\linewidth]{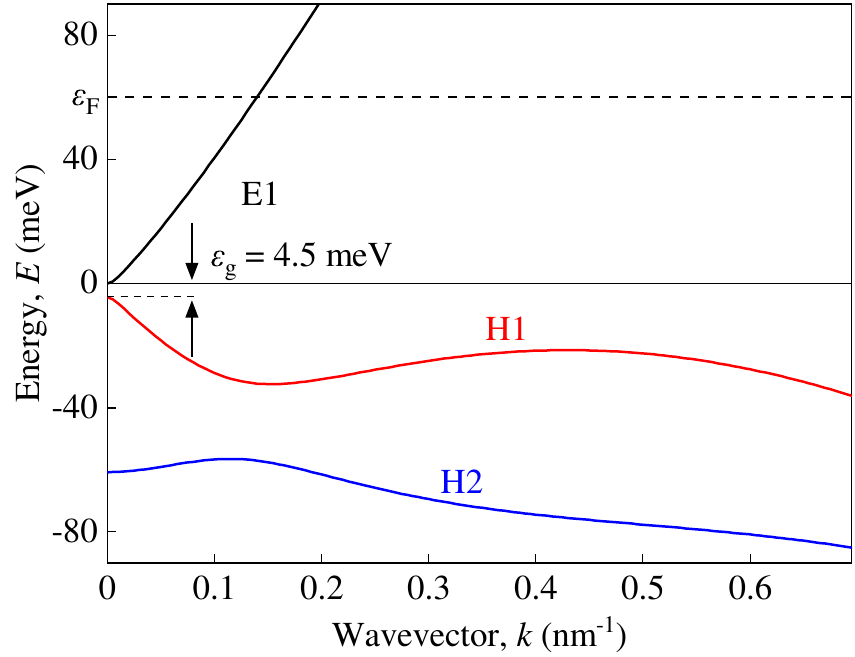}
	\caption{Calculated energy dispersion of the structures used in this study. The energy dispersion is almost linear around $k=0$ with a band gap of $\varepsilon_{\text g}=\MilliElectronVolt{4.5}$. The Fermi energy is $\varepsilon_{\text F}\approx\MilliElectronVolt{60}$.}
	\label{dispersion}
\end{figure}


\section{Discussion and theory}
\label{discussion}

Our measurements show that THz excitation of \ce{HgTe} QWs with almost linear energy dispersion leads to a photoconductivity showing a strongly nonlinear dependence on the radiation intensity. Under all conditions we observed positive photoconductivity corresponding to a decrease of the sample resistance due to illumination. 
While the fast photoconductive signal rises linearly with the radiation intensity at low intensities and small photon energies, it shows a superlinear behavior at high intensities, see Fig.~\ref{intensity}. The former signal 
is attributed to the bolometric effect caused by Drude-like absorption resulting in electron gas heating and consequently, in the change of carrier mobility. The mechanisms of this effect are well known~\cite{Ganichev2005} and are out of scope of the present paper. At high intensity the signal kinetics corresponds to the recombination time of photogenerated electron-hole pairs~\cite{Morozov2012}. The exponential growth of the signal $\Delta\sigma_{\text i}/\sigma \propto \exp(-E_0^2/E^2)$, detected at high intensities, together with its frequency dependence given by  $E_0^2\propto\omega^2$, see Fig.~\ref{field} and ~\ref{temperature},  provide an indication that it is caused by band-to-band light impact ionization \cite{Ganichev1986a,Ganichev1994,Markelz1996,Ganichev1984,Hubmann2019}. These results are in focus of our work. Below we present the theory of the impact ionization caused by the electric field of THz radiation and show that it describes well our findings.

For the theory of light impact ionization the knowledge of the band structure is crucially needed. The electron spectrum of $\NanoMeter{6.6}$ (001)-oriented \HgTe QWs is calculated  within the eight-band $\bm{k}\cdot\bm{p}$-model.
The effective Hamiltonian takes into account conduction, valence, and spin-orbit split-off bands and is taken from Ref.~\cite{Novik2005}.
\begin{figure}
	\centering
	\includegraphics[width=\linewidth]{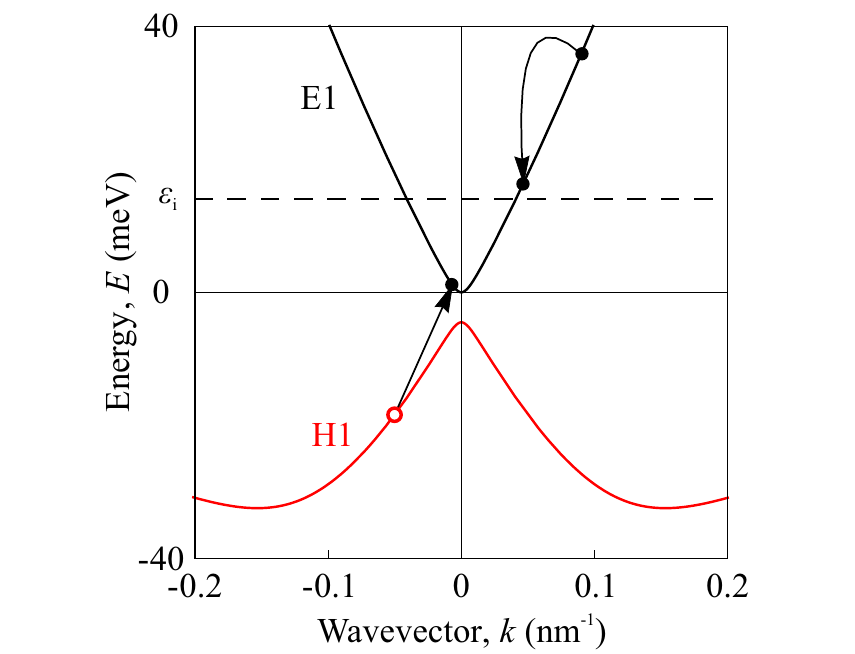}
	\caption{Sketch of ionization process. Arrows show the transitions of charge carries: An electron with high kinetic energy in the conduction band knocks the electron in the valence band by transferring part of its kinetic energy. As a result of such a process an electron-hole pair is created. }
	\label{ionization_process}
\end{figure}
Figure~\ref{dispersion} shows the electronic band dispersion calculated for the structure investigated in this work. 
The calculations show that the thickness of the QW is close to the critical width and that the band gap $\varepsilon_{\text g}=\MilliElectronVolt{4.5}$ is small. The figure reveals that the energy spectrum of electrons is close to linear, while the spectrum of holes at large wavevectors $k$ significantly differs and has a complex dependence on the wavevector. The minimum kinetic energy  $\varepsilon_i$ required for a conducting electron to lift a valence band electron into the conduction band, creating electron-hole pair is found from energy and momentum conservation for the dispersion shown in Fig.~\ref{dispersion}.
The illustration of the act of ionization is shown in Fig.~\ref{ionization_process}. Our numerical calculation shows that the threshold energy of impact ionization  $\varepsilon_i$ for this band structure is approximately equal to \MilliElectronVolt{14}, which, under our experimental conditions, is more than four times smaller than the Fermi energy $\varepsilon_F\approx \MilliElectronVolt{60}$. 
%
%
Impact ionization for $\varepsilon_{\text g} < \varepsilon_F$ was considered in~\cite{Hubmann2019} showing that under these conditions, the electron-hole pair generation is 
limited by 
the small number of free low-energy states in the conduction band. Therefore, in the ionization process heating of the electron gas is required in order to deplete occupied states in the low-energy region rather than increase the number of ``hot'' electrons.

In the following we assume that the main mechanism of electron momentum relaxation is scattering by impurities, while the energy relaxation of electrons heated by radiation is due to the interaction with optical phonons. We also suppose that the electron heating on the one hand is strong enough, so that the initially
step-like Fermi distribution function is smoothed and changes slowly on the energy scale of the optical phonon.
On the other hand, we consider that heating is not too strong for the average electron energy 
to become much higher than the initial Fermi level. These assumptions allow us to use the results of~\cite{Hubmann2019}.

For electron energies $\varepsilon$ lower than the energy of optical phonons $\varepsilon_0$, the electron distribution function is almost independent of energy and is equal to its value at $\varepsilon=\varepsilon_0$. In the region of higher energies, the distribution function is given by~\cite{Hubmann2019}
\begin{equation}
\label{distribution_f}
f_0(\varepsilon)=\dfrac{1}{1+\exp\left[- L(\varepsilon)\right]}\:, \quad L(\varepsilon)=\int^{\varepsilon_E}_{\varepsilon}
\dfrac{\varepsilon_0}{D(\varepsilon')\tau_{\text{ph}}(\varepsilon')} d\varepsilon'\:,
\end{equation}
where $\tau_{\text{ph}}^{-1}(\varepsilon)$ is the rate of electron scattering by optical phonons, $D(\varepsilon)$ is the diffusion coefficient of electrons in energy space caused by their diffusive motion in real space in the field of an electromagnetic wave and $\varepsilon_E$ is determined by the normalization by the density 
\begin{equation}
\label{normalization}
n=\int_0^{\infty}f_0(\varepsilon) g(\varepsilon) d\varepsilon.
\end{equation} 
The diffusion coefficient  $D(\varepsilon)$ has  the form
\begin{equation}
\label{diffusion_qw}
D(\varepsilon)=\dfrac{e^2 E^2 v^2(\varepsilon)}{4 \omega^2 \tau_i(\varepsilon)}\:,
\end{equation}
and the rate $\tau_{\text{ph}}^{-1}(\varepsilon)$ for the Fröhlich mechanism of electron-phonon interaction (see~\cite{Yu2010})  in quantum wells is given by
\begin{equation}
\label{tpo_linear_spectrum}
\dfrac{1}{\tau_{\text{ph}}(\varepsilon)}=
\dfrac{4 \pi^2 \varepsilon_0 e^2 g(\varepsilon)}{\overline{\epsilon}p(\varepsilon)}\:,
\end{equation}
where  $1/\overline{\epsilon}=1/{\epsilon_\infty}-1/\epsilon_0$, $\epsilon_\infty$ and $\epsilon_0$ are high- and low- frequency dielectric permittivities, $g(\varepsilon)$ is the density of states, $p(\varepsilon)$ is the electron momentum, $v(\varepsilon)$ is electron velocity and $\tau_i(\varepsilon)$ is the momentum relaxation time due to scattering by impurities. 
We note that the model assumes ${\varepsilon_i\lesssim\varepsilon_0,\varepsilon}$. 

For the studied samples the spectrum of the conduction band electrons is close to  linear, $\varepsilon=v_F p$, where $v_F$ is the Fermi velocity. Thus one obtains $g(\varepsilon)=\varepsilon/\pi\hbar^2 v_F^2$ and  $p(\varepsilon)=\varepsilon/v_F$. Hence the rate of electron scattering by phonons is given by 
\begin{equation}
\dfrac{1}{\tau_{\text{ph}}(\varepsilon)}=\dfrac{4 \pi \varepsilon_0 e^2}{v_F\overline{\epsilon}\hbar^2}\: .
\end{equation}
As addressed above, under experimental conditions, initially neutral donors are ionized by the light wave, so that electrons are scattered  by charged impurities. In the case of a linear spectrum the momentum relaxation time is given by  $\tau_i^{-1}(\varepsilon)=\tau_{iF}^{-1}(\varepsilon_F/\varepsilon)$, which for $\omega \tau_i(\varepsilon)\gg 1$ leads to 
\begin{equation}
\label{diffusion_ql}
D(\varepsilon)=\dfrac{e^2 E^2 v_F^2 \varepsilon_F}{4 \omega^2 \tau_{iF} \varepsilon}
\end{equation}
%
Using~\eqref{diffusion_qw} and \eqref{diffusion_ql} in Eq.~\eqref{distribution_f}
we obtain
\begin{equation}
\label{l_and_ee}
L(\varepsilon)=\dfrac{\varepsilon_E^2-\varepsilon^2}{\tilde{\varepsilon}^2}\:, 
\end{equation}
here we introduce the notation
\begin{equation}
\label{tildee}
\tilde{\varepsilon}^2=
\dfrac{\overline{\epsilon} v_F^3 \hbar^2 \varepsilon_F E^2}{8 \pi \varepsilon_0 \tau_{iF} \omega^2}\:.
\end{equation}
The distribution function~\eqref{distribution_f} can be written as
\begin{equation}
\label{distribution_ans}
f_0(\varepsilon)= \dfrac{1}{1+\Lambda\exp(\varepsilon^2/\tilde{\varepsilon}^2)}\:,
\end{equation}
where $\Lambda=\exp(-\varepsilon_E^2/\tilde{\varepsilon}^2)$. Taking into account both the normalization by density, Eq.~\eqref{normalization}, and distribution function, Eq.~\eqref{distribution_ans}, we obtain
\begin{equation}
\Lambda=\dfrac{1}{\exp(2\pi n\hbar^2 v_F^2/\tilde{\varepsilon}^2)-1}\:.
\end{equation}
As previously mentioned, the rate of electron-hole pair generation is proportional to the number of unoccupied states in the low-energy region $\varepsilon \ll \varepsilon_E$ of the conduction band, i.e. defined by the function ${\rho(\varepsilon)=1-f_0(\varepsilon)}$, which, according to~\eqref{distribution_ans}, is given by the expression ${\rho(\varepsilon)\approx A \exp(\varepsilon^2/\tilde{\varepsilon}^2)}$. The condition $\rho(\varepsilon)\ll 1$ at low energy region implies that $\Lambda \ll 1$, and thus
\begin{equation}
\label{function_rho}
\rho(\varepsilon)\approx \exp[-(2\pi n \hbar^2 v_F^2-\varepsilon^2)/\tilde{\varepsilon}^2]\:.
\end{equation}
The total number of excited electron-hole pairs in the sample at a given radiation intensity depends on the number of unoccupied states $\rho(\varepsilon)$ as well as on the probability of impact ionization and on the recombination rate of electron-hole pairs. The two latter quantities are unknown and as a result the exact region of energies that makes the dominant contribution to rate of generation $W$ cannot be determined. However the knowledge of this range is in fact not so important for the calculation of the functional dependence of $W$ on the radiation electric field and frequency. This is because the value of $W$ is proportional to the square of the exponential term $\rho(\varepsilon)$ (the power two arises because two electrons should be able to occupy the unoccupied states in the low-energy region). Denoting the characteristic energy corresponding to the low-energy region from \eqref{function_rho} by $\varepsilon_c$ we obtain
	\begin{equation}
	\label{generation_rate}
	W \propto \rho^2(\varepsilon_c)=\exp[-2(2\pi\hbar^2 v_F^2-\varepsilon_c^2)/\tilde{\varepsilon}^2].
	\end{equation}
Taking into account~\eqref{l_and_ee} for $\tilde{\varepsilon}$, allows us to get  the field and frequency dependencies of the number of excited electrons, which is given by
\begin{equation}
\label{generation_rate_proportionality}
W\propto \exp(-E_0^2/E^2)\:, \quad E_0^2\propto \omega^2\:.
\end{equation}

The frequency dependence in Eq.~\eqref{generation_rate_proportionality} is obtained for the relevant experimental condition  $\omega \tau_i(\varepsilon) \gg 1$. 
Generally, the dependence of the distribution function on frequency and, as a consequence, the dependence of the pair generation rate \eqref{generation_rate_proportionality} on $\omega$ are solely determined by the frequency variation of $D(\varepsilon)$ defined by \eqref{diffusion_ql}.  For an arbitrary value of $\omega \tau_i(\varepsilon)$ , according to Eq.~\eqref{diffusion_qw}, the diffusion coefficient of electrons in energy space is given by
\begin{equation}
D(\varepsilon)=\dfrac{e^2 E^2 \tau_{iF} \varepsilon/\varepsilon_F}{4(1+\omega^2 \tau_{iF}^2 \varepsilon^2/\varepsilon_F^2)}.
\end{equation}
This equation shows that for the condition $\omega \tau_i(\varepsilon) \ll 1$, the dependence of $D(\varepsilon)$ vanishes, and thus the distribution function is also independent of frequency. Carrying out the same calculation as above we again obtain an expression for the pair generation rate \eqref{generation_rate_proportionality}, however, in this case $E_0$ does not depend on the frequency $\omega$. Finally we note, that for the case of weak heating, when the above assumption of a slowly changing distribution function on the optical phonon energy scale is not fulfilled, the calculation can not be solved analytically. It can only be stated that, similarly to the case of a static electric field \cite{Keldysh1965, Dmitriev1981}, the number of excited pairs is determined by the exponent $\exp(-E_1/E)$, where $E_1$ is independent of the frequency for $\omega \tau_i(\varepsilon)\ll 1$.

Comparing the theoretical Eq.~\eqref{generation_rate_proportionality} with experiment we see that it describes the observed superlinear intensity dependence of the generation rate of electron-hole pairs at high intensities well. Indeed, experiments show that  $\Delta\sigma_{\text i}/\sigma \propto \exp(-E_0^2/E^2)$ and $ E_0^2\propto \omega^2$ agrees with Eq.~\eqref{generation_rate_proportionality}, see Figs.~\ref{field} and \ref{temperature}. The
dependence pf the photoconductive response on radiation intensity is similar to the one obtained under the same conditions for HgTe QWs with
$L_w = 5.7$~nm, Ref.~\cite{Hubmann2019}. 
Despite that the bandgaps of the structures differ by a factor of 4,
this similar behavior we attribute to
the fact that the Fermi level in both structures is much higher than the band gap
and quantum transitions, leading to impact ionization, mainly involve electrons located in the linear region of the energy spectrum.



 The observed increase of $E_0^2$ with increasing temperature indicates the reduction of the impact ionization rate, see left inset in Fig.~\ref{temperature} and is also in line with the above theory. Indeed, rising the temperature results in an increase of energy losses due to emission of phonons and, subsequently, in the reduction of radiation induced electron gas heating.  Note that the same temperature behavior was previously reported for light impact ionization of  \HgTe QWs with $L_w = \NanoMeter{5.7}$~\cite{Hubmann2019}.



While our paper is aimed to the light impact ionization we briefly discuss the saturation of the photoconductivity signal observed for the fast photoconductivity response excited by $\hbar\omega>\varepsilon_g$, see Fig.~\ref{highfrequency}. In this case, 
radiation absorption due to inter-band optical transitions is also possible, in addition to Drude-like transition. At low intensities the inter-band transitions play almost no role because the final states of these transitions are lying below the Fermi energy and thus, these states are occupied. At high intensities, however, electron gas heating, discussed above, depletes occupied states in the low-energy region of the conduction band and direct band-to-band transitions contribute to the photoconductivity signal. The interplay of the signal components caused by Drude absorption, direct band-to-band transitions and impact ionization causes a complex intensity dependence of the total signal, see Eq.(\ref{highfreqfit}). 
Two first terms on the right hand side of this equation we attribute to direct optical transitions and Drude absorption. The saturation of the fundamental absorption at high intensities is a well known process. Recently, it has been demonstrated that Drude-like transitions under strong electron gas heating may also saturate with intensity increase~\cite{Mics2015}.  A more detailed analysis of these processes is out of scope of this paper. The last term in Eq.(\ref{highfreqfit}) describes the impact ionization. Due to the strong frequency dependence of this process, see Fig.~\ref{field} and Eq.~\eqref{generation_rate_proportionality}, for $\hbar\omega>\varepsilon_g$ its contribution is clearly detected for the lowest frequency only, see Fig.~\ref{highfrequency}. 

\section{Summary} 
\label{summary}

To sum up our work, by studying \HgTe QWs with nearly linear energy dispersion we observed that high power THz radiation results in band-to-band impact ionization. The developed theory, considering the impact ionization for arbitrary values of  $\omega \tau_i(\varepsilon)$, describes the experimental findings well. It shows that in our experiments the light-impact ionization by the electric field of THz radiation is realized. In this case $\omega \tau_i(\varepsilon)\gg 1$ and the ionization rate depends drastically on the radiation frequency. 

\section{Acknowledgments}
\label{acknow}
We thank E.L. Ivchenko for helpful discussions. The support from the Deutsche Forschungsgemeinschaft (DFG, German Research Foundation) – Project-ID 314695032 – SFB 1277,
the Elite Network of Bavaria (K-NW-2013-247) and the Volkswagen Stiftung Program (97738) is gratefully acknowledged. G.V. Budkin acknowledges the support of “BASIS” foundation. D.W. acknowledges funding from the European Research Council (ERC Advanced Grant, Nr. 787515-ProMotion)

\end{document}